\documentclass{elsart}

 \usepackage{graphicx}
 \usepackage{epsfig}

\usepackage{amssymb}
\def\pom{{I\!\!P}}
\begin{document}

\begin{frontmatter}



\title{Hard and Soft Contributions in Diffraction\thanksref{FFK}}
\thanks[FFK]{Based on the talk presented by M.B. Gay Ducati at the IXth Blois
Workshop, Pruhonice near Prague, June 9-15, 2001.} 

\author{M.B. Gay Ducati $^a$, V.P. Gon\c{c}alves $^b$, M.V.T. Machado $^a$}

\address{$^a$ Instituto de F\'{\i}sica, Universidade Federal do Rio Grande do
Sul \\  Caixa Postal 15051, 91501-970 Porto Alegre, RS, BRAZIL.\\
$^b$ Instituto de F\'{\i}sica e Matem\'atica, Universidade
Federal de Pelotas\\
Caixa Postal 354, CEP 96010-090, Pelotas, RS, BRAZIL
 }

\begin{abstract}
We report our calculations on the logarithmic slope of the
diffractive structure function $F_2^D$ considering  distinct approaches
coming from both the pQCD and the Regge descriptions of diffractive DIS. Such a
quantity is a potential observable to discriminate between pQCD and Regge
domains and allows to disentangle the underlying dynamics.  \end{abstract}

\end{frontmatter}

\section{Introduction}

The experimental measurements of the logarithmic slope ($Q^2$-slope) of the
inclusive structure function $F_2(x,Q^2)$ presented a new challenge for the
current small $x$ approaches \cite{slopedata}. The standard DGLAP
\cite{DGLAP} dynamics was found to fail in describing the  low
momentum transfer $Q^2<1$ GeV$^2$ region in the correlated ($x$, $Q^2$)
plane, overestimating the data. The QCD expected result is that the slope
should be proportional to the nucleon gluon distribution. Thus, new phenomena
could be present modifying the gluonic content of the proton or a
transition between hard and soft domains would be emerging. 
Indeed, the
earlier observed turn over in the $Q^2$-slope was considered as  an evident
signal for change of the dynamics (see e.g., Ref. \cite{slopephen}). The most
recent non-correlated measurements \cite{slopenew} support a broad DGLAP
description using the updated pdf's. However, the turn over observed when
one considers the slope at fixed center of mass energy $W$ still  deserves
criterious work \cite{Thorne2001}. Therefore, a complete  description of
the $F_2(x,Q^2)$ logarithmic $Q^2$-slope in the full kinematical range is 
not known at present and suggests an interplay or a competition between hard
and soft regimes.

When we look into diffraction dissociation in deep inelastic scattering, in
particular the diffractive structure function $F_2^D$, an interplay between
hard and soft domains is more explicit \cite{Mueller}. The soft content in
diffractive dissociation is known to be large, justifying the extensive
non-perturbative phenomenology used to describe its energy dependence.
Regarding this feature, the structure function $F_2^D$ is a good test for 
Pomeron models, and the question if we have in the real world  a
perturbative (BFKL) or  a soft Pomeron is  an open  discussion in the
community. The main advantage from the Regge inspired models is a quite simple
phenomenological description of the process. However, they  cannot constrain
in a theoretical ground the behavior on $Q^2$ and $\beta$ variables. On the
other hand, a great progress has been made in perturbative QCD, setting the
correspondent dominant contributions order by order in perturbation theory.
The notions of photon wavefunction and dipole cross section are now the basic
blocks to describe perturbatively the diffractive events. The main appeal of
the pQCD approaches is to produce very clear predictions concerning the
$\beta$  spectrum, as well as setting a deeper  connection with the  machinery
builded for the inclusive case (DIS).

In a similar way as the $F_2$ slope case, the logarithmic slope of the
diffractive structure function $F_2^D$ should be studied and its role into
disentangle dynamics needs investigation. The experimental status concerning
$F_2^D$ provides high precision data in a wide kinematical region,
possibly allowing to extract the slope with good statistic precision. Bearing
in mind these issues, we have proposed \cite{slplb,slnpa} to calculate this
quantity from the available models describing diffractive DIS. In particular,
we choose two representative approaches in order to obtain quantitative
results. One of them is the pQCD approach \cite{pQCD}, where the
diffractive process is modeled as the scattering of the photon Fock states
with the proton through a gluon ladder exchange (in the proton rest frame).
Moreover, we consider a Regge inspired model \cite{CKMT}, where the
diffractive production is dominated by the soft Pomeron. These approaches
differs basically in the Pomeron exchange definition and its coupling with the
incident photon.

In Ref. \cite{slplb} we have 
calculated the slope for both pQCD and Regge approaches considering a
kinematical constraint for the correlated variables $x$ and $Q^2$. In Ref.
\cite{slnpa} such constraint was disregarded, and the calculations were also 
extended to include the analysis of the saturation model \cite{GBW}.
However, here we report basically the conclusions presented in \cite{slplb}.
In a general pointview, the distinct behaviors obtained for the logarithmic
slope from different physical approaches could give hints in the underlying
dynamics and which can be tested if such  observable is to be experimentally
measured. 

\section{Analytical results for the Regge based model}

A few years ago Capella-Kaidalov-Merino-Tran Thanh Van (CKMT)
proposed a model for diffractive DIS based on Regge theory 
and the Ingelman-Schlein ansatz \cite{CKMT}. The Pomeron is considered as a Regge pole with a
trajectory $\alpha_{\pom}(t)=\alpha_{\pom}(0) + \alpha^{\prime}_{\pom}\, t $
determined from soft processes, in which absorptive corrections (Regge cuts)
are taken into account. Explicitly, $\alpha_{\pom}=1.13$ and $%
\alpha^{\prime}_{\pom}=0.25\;$ GeV$^{-2}$. The diffractive contribution to DIS
is written in the factorized form:
\begin{eqnarray}  \label{flux}
F_2^D(x_{\pom},\beta, Q^2, t)=\frac{[g^{\pom}_{pp}(t)]^2}{16 \pi}x_{\pom%
}^{1-2\alpha_{\pom}(t)} \, F_{\pom}(\beta,Q^2,t)\;,
\end{eqnarray}
where the details about the coefficients appearing in the Pomeron
flux can be found in Refs. \cite{CKMT}. One of the main points of this model
is the dependence on $Q^2$ of the Pomeron intercept [$F_{\pom} \propto
\beta^{\Delta (Q^2)}$]. For low values of virtuality (large cuts), $\Delta$ is
close to the effective value
found from analyzes of the hadronic total cross sections ($\Delta \sim 0.08$%
), while for high values of $Q^2$ (small cuts), $\Delta$ takes the
bare Pomeron value, $\Delta \sim 0.2-0.25$. The comparison of the CKMT model
with data is quite satisfactory, mainly when considering a perturbative
evolution of the Pomeron structure function. We notice that here one uses the
pure CKMT model rather than including QCD evolution of the
initial conditions , which has been considered to
improve the model at higher $Q^2$. Such procedure ensures that we
take just a  strict Regge model, namely avoiding contamination by pQCD
phenomenology.

Justifying our  choice, the CKMT  is a long standing approach
describing in a consistent way  both inclusive and diffractive
deep inelastic based on Regge theory, which is continuously
improved  considering updated experimental results
\cite{Kaidepjc}.  A subtle
question in CKMT approach is the dependence of the Pomeron
intercept on the virtuality for the inclusive case:  the smooth interpolation
between a soft and a semihard intercept seems to break down the pure reggeonic
feature of the model.  However, in diffractive DIS the energy
dependence (i.e., the Pomeron flux) is driven by a soft Pomeron with a fixed 
$\alpha_{\pom}=1.13$, properly corrected by absorptive effects. Indeed, this
fact is verified in \cite{slplb}, when considering the
effective slope $\partial \ln F_2^D/ \partial \ln (1/x_{\pom})$. 

Considering all properties of the diffractive structure function
in the CKMT model, the calculation of the logarithmic slope is
straightforward. The expression reads now:
\begin{eqnarray}
 \frac{\partial F_2^D}{\partial \,\ln Q^2} & = &  N \,\,
x_{\pom}^{1-2\,\alpha_{\pom}(0)}\, \left[ \, eA \, \, \beta^{-
\Delta(Q^2)}\,(1 - \beta)^{n(Q^2)\,+\,2}\, \left( {\frac{Q^2 }{Q^2
+ a}} \right )^{1\, +\, \Delta(Q^2)}\,S_{\pom}(Q^2,\beta)\,
 \,\,  \,\, \right.  \nonumber \\
& + &  \, \left.  \,fB \, \beta^{1 - \alpha_R} \, (1 -
\beta)^{n(Q^2)\,-\,2} \,\left( {\frac{Q^2 }{Q^2 + b}}
\right)^{\alpha_R} \, S_R(Q^2,\beta)\, \right] \,\,,
\end{eqnarray}
where the overall normalization $ N$ comes from
the integration over $t$ of the Pomeron flux, $n(Q^2) = {\frac{3
}{2}} \left ( 1 + {\frac{Q^2 }{Q^2 + c}} \right )$, $\alpha_R$ is
the  secondary reggeon intercept (only the $f$ meson).  The coefficients and
constants are taken from \cite{CKMT}, and the factors $S_{{\pom}%
,R}(Q^2,\beta)$ are defined as
\begin{eqnarray}
S_{\pom}(Q^2, \beta) & = & \Delta(Q^2)\,\left[\, \frac{a}{Q^2 + a}\,
\right]\,+\,\frac{3\,c}{2}\,\left[\, \frac{Q^2}{(Q^2+c)^2}%
\,\right]\,\ln(1-\beta) \, +   \nonumber \\
& + & \, 2\,d\,\Delta_0\,\left[\,\frac{Q^2}{(Q^2 + d)^2} \,\right] \, \ln
\left( \frac{Q^2}{\beta\,(Q^2 + a)} \right) \,\,, \label{sp} \\
S_R(Q^2,\beta) & = & \,\, \alpha_R(0) \, \left[ \, \frac{b}{Q^2 + b} \,
\right] \, + \, \frac{3\,c}{2}\,\left[ \, \frac{Q^2}{(Q^2+c)^2}
\,\right]\,\ln(1-\beta) \,\,. \label{sr}
\end{eqnarray}

\section{Analytical results for the pQCD based model}

Although the diffractive dissociation to be mainly connected to soft
processes and thus linked with the Regge theory, the pQCD framework
has been recently used  to describe quite well the diffractive
structure function. In particular, we have considered  the sound model in
\cite{pQCD}. The physical picture is that, in the proton rest frame,
diffractive DIS is described by the interaction of the photon Fock states
($q\bar{q}$ and $q\bar{q}g$ configurations) with the proton through a Pomeron
exchange, modeled as a two hard gluon exchange. The corresponding structure
function contains the contribution of $q\bar{q}$ production to
both the longitudinal and the transverse polarization of the
incoming photon and of the production of $q\bar{q}g$ final states
from transverse photons. 

The basic elements of this approach are
the photon light-cone wavefunction and the unintegrated gluon
distribution (or dipole cross section in the dipole formalism).
For elementary quark-antiquark final state, the wavefunctions
depend on the helicities of the photon and of the (anti)quark. For
the $q\bar{q}g$ system one considers a gluon dipole, where the
$q\bar{q}$ pair forms an effective gluon state associated in color
to the emitted gluon and only the transverse photon polarization
is important. The interaction with the proton target is modeled
by two gluon exchange, where they couple in all possible
combinations to the dipole. In a comparison with data, the
transverse $q\bar{q}$, $q\bar{q}g$ production
and the longitudinal $q\bar{q}$ production dominate in distinct regions in $%
\beta$, namely medium, small and large $\beta$ respectively \cite{pQCD}. The $%
\beta$ spectrum and the $Q^2$-scaling behavior follow from the
evolution of the final state partons, and they are derived from the
light-cone wavefunctions of the incoming photon, decoupling from
the dynamics inside the Pomeron. The energy dependence and
the normalizations are free parameters.

The calculations  for the $F_2^D$-logarithmic
slope are simple to perform, considering each contribution coming
from the different configurations of the photon Fock state. Moreover, this approach allows to obtain
parameter free predictions for the logarithmic slope, since all
coefficients have been obtained in comparison with the H1 and ZEUS
data. We justify our choice due to the simple analytic expressions for the
diffractive structure function regarding each Fock state configuration,
which turns the analyzes clearest and avoid cumbersome
numeric calculations. The expressions for the distinct
contributions to $F_2^{D(3)}$  as
well as the remaining parameters can be found in Ref. \cite{pQCD}. Thus, the
diffractive logarithmic slope is written as \begin{eqnarray}
\frac{d\,F_2^{D(3),\,q\bar{q}T}(x_{\pom}, \beta, Q^2)}{d\,\ln Q^2} &=&
\frac{n^1_2}{ (\, \ln \frac{Q^2}{Q^2_0} +
1\,)}\,\,F_2^{D(3),\,q\bar{q}T}(x_{\pom}, \beta, Q^2) \,\,, \label{dpqcd} \\
\frac{d\,F_2^{D(3),\,q\bar{q}G}(x_{\pom}, \beta, Q^2)}{d\,\ln Q^2}
&=& \frac{1}{(\, \ln \frac{Q^2}{Q^2_0} + 1\,)}\,\left[\, n^1_2
\,+\, \frac{Q^2}{Q^2+Q^2_0}
\,\right]\,\,F_2^{D(3),\,q\bar{q}G}(x_{\pom}, \beta, Q^2) \,\,,
 \nonumber \\
\frac{d\,F_2^{D(3),\,q\bar{q}L}(x_{\pom}, \beta, Q^2)}{d\,\ln Q^2}
&=& \left[ \frac{n^1_4}{(\,\ln \frac{Q^2}{Q^2_0} + 1\,)} \,+\,
\frac{Q^2 - 7\,\beta Q^2_0}{Q^2 + 7\,\beta Q^2_0}
\,\right]\,\,F_2^{D(3),\,q\bar{q}L}(x_{\pom}, \beta, Q^2) \,\,.
\nonumber
\end{eqnarray}

\section{Results and Discussions}

\begin{figure}[t]
\centerline{\psfig{file=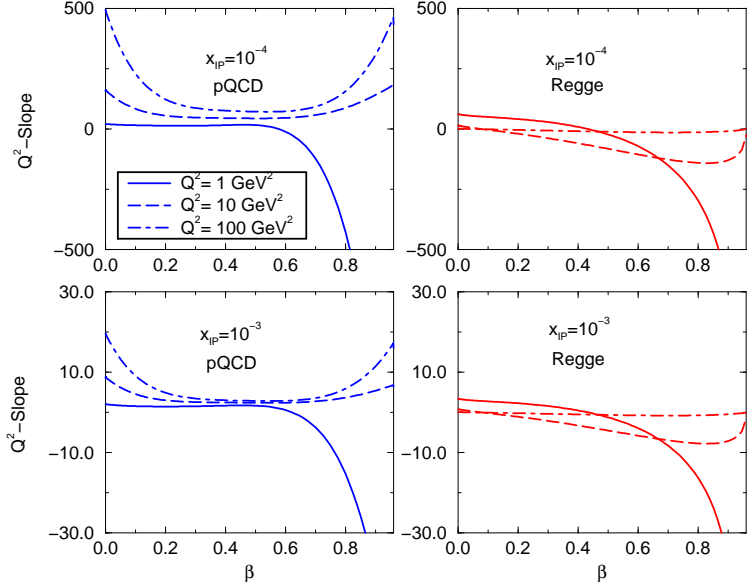,width=90mm}}
\caption{The $Q^2$-slope versus
$\beta$ for the pQCD (on the left) and CKMT (on the right) models. }
\label{fig5}
\end{figure}

\begin{figure}[t]
\centerline{\psfig{file=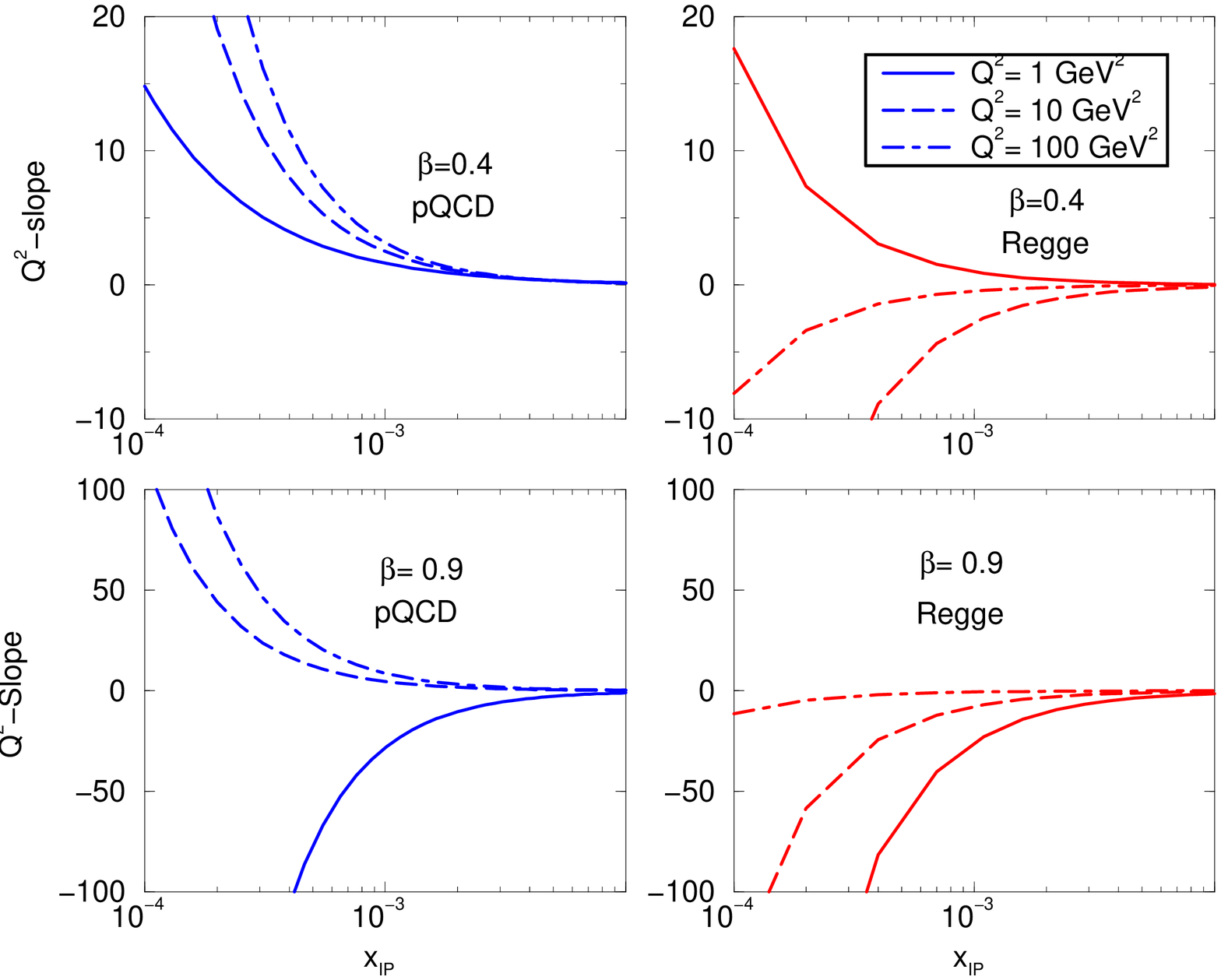,width=90mm}} 
\caption{The $Q^2$-slope
versus $x_{\pom}$ for the pQCD (on the left) and CKMT (on the
right) models. }
\label{fig6} 
\end{figure}

In Figs. (\ref{fig5})-(\ref{fig6}) we present a comparison between the 
models mentioned above. In Fig. (\ref{fig5})
one shows the $\beta $-dependence for typical values of $x_{\pom}$ and $Q^2$%
, where the momentum transfer is ranging from 1 up to 100 GeV$^2$. 
The CKMT
model predicts a flat behavior on the whole $\beta $ range. Particularly, at %
$Q^2=1$ GeV$^2$ there is a strong decreasing of the slope at large $\beta $%
. This is due to the presence of factors $\ln (1-\beta )$ in the second
terms of Eqs. (3)-(4). For larger $Q^2$ values, the logarithm on %
$Q^2$ appearing at the third term of Eq. (\ref{sp}), compensates the
decreasing. At 100 GeV$^2$ of momentum transfer, the CKMT
predicts a flat behavior  of the slope for the whole  $\beta $ spectrum. The
pQCD model produces an increasing of the slope in both small and large values
of $\beta $, while presents a flat behavior at the medium one. This
increasing of the slope is due to the enhancements in the $q\bar{q}G$
(dominant at low $\beta $) and $q\bar{q}L $ (leading at high $\beta $)
contributions from the $Q^2$-factors appearing in Eqs. (5), respectively. A
steep $Q^2$-slope decreasing into negative values is also present at the
virtuality $ Q^2=1$ GeV$^2$ for large $\beta $, in a similar way to the CKMT
model. However, beyond the contribution of the (dominant) longitudinallly
polarized $q\bar{q}$ pair configuration, this region also receives
contributions associated to the $q\bar{q}G$ configuration, as discussed in
\cite{slnpa}. 

In Fig. (\ref{fig6}) one presents the $x_{\pom}$ dependence at medium and
large $\beta $. The low $\beta $ region was disregarded in order to avoid to
deal with subleading reggeonic contributions, important in this kinematical
domain. We observe again a smooth behavior for the pQCD predictions, with a
positive slope in almost the whole range (negative values are present at
both large $\beta $ and $Q^2=1\,GeV^2$, in agreement with the discussion in
the previous paragraph). In Ref. \cite{slnpa} we have found that the saturation
model produces a transition between positive and negative slope values at low
$\beta=0.04$, while shows a positive slope for medium and large $\beta$,
differing from the results for the pQCD model discussed above. Since  the
diffractive cross section is strongly sensitive to the infrared cutoff, one of
the main differences between these two models is the assumption related to the
small $Q^2$ region. Returning, the CKMT model predicts predominantly negative
slope values in this kinematical domain, converging to a flat value for larger
$x_{\pom}$. A strict difference between the predictions of the two models is
the change of signal of the slope with the $Q^2$ evolution at medium $\beta $,
present in the CKMT results. 

\begin{figure}[t]
\centerline{\psfig{file=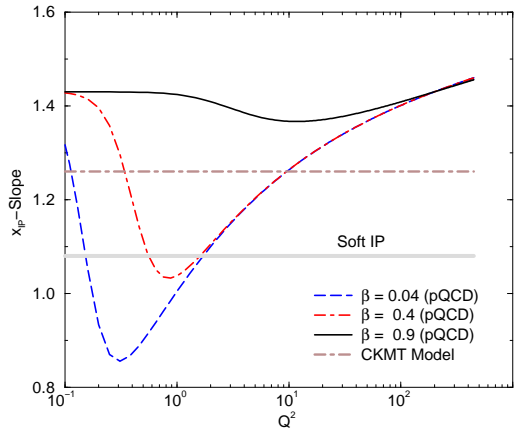,width=80mm}}
\caption{The $x_{\pom}$-slope as a function of the photon virtuality $Q^2$.}
 \label{fig3}
\end{figure}

In  Fig. (\ref{fig3}) we show the results for $d \ln F_2^D/d
\ln(1/x_{\pom})$ (or shortly, $x_{\pom}$-slope) as a function of the photon virtuality $Q^2$. 
Indeed, this
quantity gives the Pomeron intercept and its behavior describes the energy dependence of the 
diffractive structure function. While the CKMT model predicts a
constant value, without dependence on  $\beta$, the pQCD model is
$\beta$-dependent. This
feature is  associated to  the  distinct energy  dependence of each  term in
Eqs. (5), which dominate at specific regions of the phase space. For instance,
a hard intercept is observed from the slope at large $\beta = 0.9$. In fact, a
dependence on $\beta$ for the Pomeron intercept is expected as shown in Ref.
\cite{pQCD}. We notice that the pQCD model is only  valid above the  starting
point $Q_0^2=1$ GeV$^2$ and our extrapolation for lower values is for sake of
comparison. For completeness we include the soft Pomeron intercept in the
plot. We verify, therefore, the evident distinction between the predictions
from the CKMT and pQCD based approaches.

In summary, the results above  allow to discriminate the behaviors
predicted from the different approaches, namely perturbative QCD (hard
physics) and non-perturbative (soft) physics. Theoretically, the difference
between  comes from the ansatz for the photon-proton interaction (hard or soft)
and also from the relation of the Pomeron structure function  with the proton 
inclusive one, for instance  considered in the CKMT model. Such a  relation
implies the inclusion at most of the $q\overline{q}_T$ 
configuration in the photon wavefunction in the CKMT estimates. On the other
hand, the pQCD model analyzed here includes the contribution of gluon emission
in the photon wavefunction, thus taking into account the leading twist
contributions $q\overline{q}_T$  and  $(q\overline{q}g)_T$. Moreover, the
higher twist piece, $q\overline{q}_L$ is also included in the description.
Therefore, the analyzes of the $\beta $ spectrum of the $F_2^D$-slope would be
important in an experimental study. Moreover, concerning  the $x_{\pom}$
spectrum, the signal of the slope at medium values of $Q^2$ ($Q^2\approx 10$
GeV$^2$) is a good source of information about the dynamics.

\end{document}